\documentclass[letter]{aa}

\usepackage{graphicx}
\usepackage[dvips]{color}
\usepackage[pdftex]{hyperref}
\usepackage{txfonts}
\usepackage{natbib}

\bibpunct{(}{)}{;}{a}{}{,}
\begin{document}
   \title{The metallicity of gamma-ray burst environments from high energy
          observations}

   \subtitle{}

   \titlerunning{Metallicity of GRB environments using $\gamma$-rays}

   \author{D.~Watson
          \inst{1}
          \and
          P.~Laursen\inst{2}
          }

   \authorrunning{Watson \& Laursen}

   \institute{Dark Cosmology Centre, Niels Bohr Institute, University of
              Copenhagen, Juliane Maries Vej 30, DK-2100 Copenhagen \O, Denmark\\
              \email{darach@dark-cosmology.dk}
              \and
              Oskar Klein Centre, Dept.~of Astronomy, Stockholm University,
              AlbaNova, SE-10691 Stockholm, Sweden\\
              \email{plaur@astro.su.se}
             }

   \date{Received XXX; accepted XXX}

  \abstract
  {Gamma-ray bursts (GRBs) and their
   early afterglows ionise their circumburst material. Only
   high-energy spectroscopy therefore, allows examination of the matter close to the
   burst itself.  Soft X-ray absorption allows an estimate to be made of the
   total column density in metals. The detection
   of the X-ray afterglow can also be used to place a limit on the total gas
   column along the line of sight based on the Compton scattering opacity. 
   Such a limit would enable, for the first time, the determination of lower
   limits on the metallicity in the circumburst environments of GRBs.}
  {In this paper, we determine the limits that can be placed on the
   total gas column density in the vicinities of GRBs based on the Compton
   scattering.}
  {We simulate the effects
   of Compton scattering on a collimated beam of high energy photons passing
   through a shell of high column density material to determine the expected
   lightcurves, luminosities, and spectra.  We compare these predictions to
   observations, and determine what limits can realistically be placed on
   the total gas column density.}
  {The smearing out of pulses in the lightcurve from Compton scattering is
   not likely to be observable, and its absence does not place strong
   constraints on the Compton depth for GRBs.  However, the distribution of
   observed luminosities of bursts allows us to place statistical,
   model-dependent limits that are typically $\lesssim10^{25}$\,cm$^{-2}$
   for less luminous bursts, and as low as $\sim10^{24}$\,cm$^{-2}$ for the
   most luminous.  Using the shape of the high-energy broadband spectrum,
   however, in some favourable cases, limits as low as
   $\sim5\times10^{24}$\,cm$^{-2}$ can placed on individual bursts,
   implying metallicity lower limits from X- and gamma-rays alone from 0 up
   to $0.01\,Z/Z_\odot$.  At extremely high
   redshifts, this limit would be at least $0.02\,Z/Z_\odot$, enough to
   discriminate population III from non-primordial GRBs.}
   {}


   \maketitle

\section{Introduction}
 
 A major issue in modern cosmology is our understanding of the
 metal-enrichment history of the universe: how, when and where the
 non-primordial elements were synthesised.  Most avenues open to us to
 investigate this question at high redshifts rely on bright sources such as
 active galactic nuclei and gamma-ray bursts (GRBs) as back-lighting
 for absorption spectroscopy of distant galaxies.  Gamma-ray burst
 afterglows are bright enough to examine the contents of star-forming
 galaxies by absorption spectroscopy.  Most detailed information so far has
 come from optical and ultraviolet (UV) observations.  Due to the strong
 ionising effect of the burst and early afterglow on the matter surrounding
 a GRB however
 only high-energy spectroscopy can probe the matter close to the burst. 
 X-ray spectra allow examination of this matter, which we now know lies
 within only a few parsecs \citep{2007ApJ...660L.101W} and is ionised to a
 fairly high degree \citep{2008ApJ...685..344P,2010arXiv1010.2034S} and is
 therefore invisible at optical and UV wavelengths.

 Photoelectric absorption by inner shell electrons absorb X-rays and this
 allows an estimate to be made of the total column density in metals.  The
 total gas column is unavailable, however, since the hydrogen, which is
 ionised, is transparent to the X-rays even at moderate column densities. 
 However, at extreme column densities, ionised hydrogen is no longer
 transparent to high energy photons since X-ray scattering off the free
 electrons has a cross section approximated by the Thomson cross section at
 low energies, $\sigma_{\mathrm{T}}\sim6.65\times10^{-25}$\,cm$^2$.  This
 fact has been used to derive limits on the column density of ionised
 hydrogen in the immediate surroundings of the GRB\,050904
 \citep{2007ApJ...654L..17C}, and to exclude Comptonisation as the origin of
 very hard gamma-ray emission in some GRBs \citep{2003A&A...406..879G}.  It
 has also recently been suggested by \citet{2010arXiv1008.3054C} that the
 detection of the X-ray and gamma-ray emission from GRBs could be used to
 place a limit on the total gas column along the line of sight based on the
 fact that at very high column densities Compton scattering will essentially
 eliminate all of the X- and gamma-ray emission and that this could be
 useful for a future GRB mission.  Such a limit would allow us to estimate
 lower limits to the metallicity in the immediate environments of GRBs by
 providing an upper limit to the total gas column density.  At high
 redshifts, not only could this allow us to determine metallicities without
 the problems associated with UV spectra, but it would also allow us to
 determine the metal abundances right in the hearts of star-forming regions.

 Future X-ray missions equipped with large effective area, high resolution
 spectrographs, would allow redshifts to be determined directly from X-rays
 without reference to optical or UV light, enabling far more complete and
 unbiased estimates of GRB environments to be obtained.  However, without
 this technique, they could never determine the metallicities of the regions
 they probe.  In placing this Compton-thick limit, however
 \citet{2010arXiv1008.3054C} assume for simplicity that this limiting column
 density is approximately one optical depth $\tau=1/\sigma_{\mathrm{T}}$. 
 But given that the observed column densities of GRBs span a range from
 $<10^{21}$\,cm$^{-2}$ up to $\sim10^{23}$\,cm$^{-2}$
 \citep{2009MNRAS.397.1177E,2010MNRAS.402.2429C}, the value of the limit
 that can be placed on the total column density makes a substantial
 difference to how interesting the metallicity lower limit obtained will be. 
 In this paper we attempt to quantify, via Monte Carlo simulations, the
 limits that can placed on the total column density and hence metallicities
 of the circumburst medium from the Compton-transparency of GRBs and their
 afterglows.

\section{Simulations}

\subsection{Burst modeling}
\label{sec:burst}

The radiative transfer (RT) calculations are conducted using a modified
version of the Monte Carlo RT code {\sc MoCaLaTA}, originally designed for
simulating the scattering of Ly$\alpha$ photons in the interstellar and
intergalactic media \citep{2009ApJ...696..853L}.
Although the code is capable of assuming an arbitrary distribution of gas
density, temperature, velocity field, etc., taken e.g.\ from a hydrodynamic
simulation, for the purpose of the present calculations the burst is modeled as
a central source emitting a power-law spectrum of photons with energies $E$
between 0.1 keV and 1 MeV and a photon spectral index of $\Gamma = 2$.
That is, the emitted spectrum is given by
\begin{equation}
\label{eq:Fem}
F_{\mathrm{em}}(E) = A\,E^{-\Gamma},
\end{equation}
where $A$ is a constant.
The source is surrounded by a thick, spherical shell of electron
column density $N_e$, with a temperature $T$ and velocity field
$\vec{v}_{\mathrm{bulk}}$. Photons are emitted in two narrow cones of
opening angle $\vartheta = 5^\circ$, taken to lie along the $z$ axis.

\subsection{Radiative transfer}
\label{sec:RT}

While at low energies a photon interacting with an electron is scattered in a
more or less random direction, at high energies the phase function is
characterised by a significant probability of forward scattering.
Moreover, the cross section $\sigma_{\mathrm{C}}$ decreases with increasing
photon energy. These effects are given by the differential cross section
\citep{1929ZPhy...52..853K}
\begin{equation}
\label{eq:KN}
\frac{d\sigma_{\mathrm{C}}}{d\Omega} = \frac{1}{2} r_e^2 P^2(E,\theta)
   \left(
      P(E,\theta) + \frac{1}{P(E,\theta)} - 1 + \cos^2\theta,
   \right)
\end{equation}
where $r_e$ is the classical electron radius, and
\begin{equation}
\label{eq:P}
P(E,\theta) = \frac{1}{1 - \alpha(1 - \cos\theta)}
\end{equation}
is the energy ratio of the incident and the scattered photon, with
$\alpha \equiv E/m_e c^2$ being the photon energy in terms of the electron
rest energy.

The photons are followed as they scatter stochastically through the medium
(or escape freely).
Equations \ref{eq:KN} and \ref{eq:P} are expressed in the reference frame of
the electron. Thus, at each scattering the energy of a photon is first Lorentz
transformed to the reference frame of the electron, given by the sum of
$\vec{v}_{\mathrm{bulk}}$ and a thermal velocity drawn from a Gaussian
distribution.
At high energies the change in energy is dominated by the Compton effect
described by Eq.~\ref{eq:P}; however, at lower energies the Doppler shift induced by
high velocity electrons may become important.

{\sc MoCaLaTA} has already been tested extensively against various
analytical solutions; the additional subroutines implemented in this study
were tested correspondingly.  Furthermore, a variety of temperatures,
velocity fields, density fields, and opening angles were tested.  However,
as will be shown in Section Sect.~\ref{sec:lc}, observationally only the
prompt emission is of interest, which depends almost exclusively on the
column of intervening gas and not on its actual structure

\section{Results}

\subsection{Lightcurves}
\label{sec:lc}

\begin{figure}
\includegraphics[width=\columnwidth,clip=]{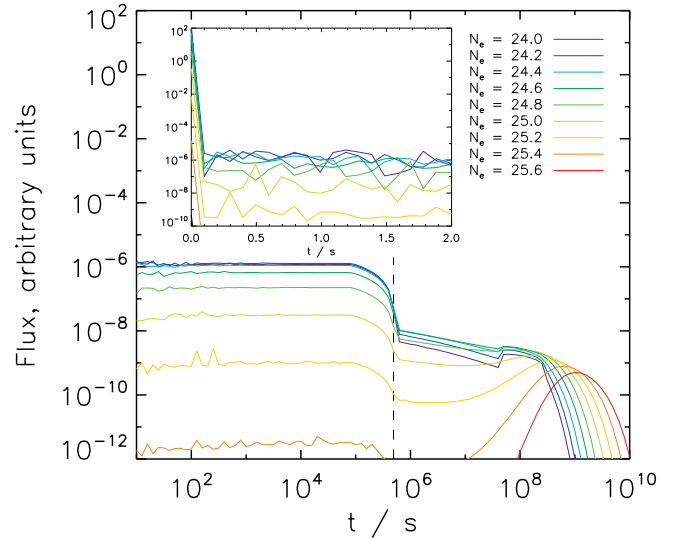}
\caption{Lightcurves for a range of electron column densities $N_e$. Only the
         prompt emission will be detectable with current instrumentation for
         cosmological GRBs.  The inset is a zoom-in of the prompt emission. 
         The vertical \emph{dashed} line marks the maximum time lag for
         singly scattered photons, as given by Equation~\ref{eq:Dt}.  }
\label{fig:lightcurve}
\end{figure}

Sampling the photons according to the time it takes to reach the observer,
a lightcurve can be obtained, giving the photon flux as a function of time.
Fig.~\ref{fig:lightcurve} shows such lightcurves for a range of electron column
densities.
The curves are characterised by three parts of different physical origin:
The initial peak---the ``prompt'' emission---consists of photons that escape
the medium directly, without scattering, and hence its height is simply
specified by the optical depth along the line of sight to the burst.  The
following intermediate phase consists of photons that are scattered once,
and thus its extension is given by the size of the cloud and
the opening angle of the jet; for a spherical cloud of radius $R = 1.25$\,pc
and opening angle $\vartheta = 5^\circ$, this phase lasts
\begin{equation}
\label{eq:Dt}
\Delta t = \frac{R}{c} (1 - \cos\vartheta) \simeq 5^{\rm d}16^{\rm h}.
\end{equation}
Finally, there is an afterglow of photons
that scatter several times and thus perform a random walk out of the cloud. 
Since the number of steps in a random walk scales with $N_e^2$, but the step
size decreases linearly with $N_e$ for a fixed physical size of the cloud,
the maximum of the afterglow approaches a time proportional to $N_e$.

For high values of column density, the above considerations become somewhat
inaccurate, since at each scattering, a photon---especially a high-energy
photon---loses energy, thus resulting in a slowly decreasing electron cross
section.

Comparing the magnitude of the observed simulated prompt peak with the
energy emitted by the burst, constraints can be put on the Compton thickness of the
circumburst material. For the emitted spectrum in Equation\ref{eq:Fem},
the observed spectrum of the prompt emission will be
\begin{equation}
\label{eq:Fobs}
F_{\mathrm{obs}}(E) = A\,E^{-\Gamma}\,e^{-N_e\sigma_{\mathrm{C}}(E)}.
\end{equation}

\begin{figure}
\includegraphics[width=\columnwidth,clip=]{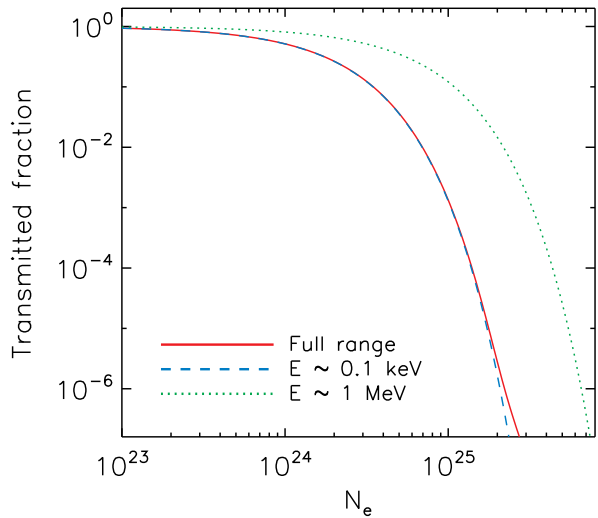}
\caption{Fraction of emitted photons that are transmitted directly through a
         cloud of column density $N_e$ for the full energy range (\emph{red
         solid}), low-energy photons (\emph{blue dashed}), and high-energy
         photons (\emph{green dotted}).}
\label{fig:f}
\end{figure}

The fraction $f$ of the flux observed as prompt emission is thus
\begin{equation}
\label{eq:f}
f = \frac{\int_{E_{\mathrm{min}}}^{E_{\mathrm{max}}}
          E^{-\Gamma}\,e^{-N_e\sigma_{\mathrm{C}}(E)}\,dE}
         {\int_{E_{\mathrm{min}}}^{E_{\mathrm{max}}}
          E^{-\Gamma}\,dE}.
\end{equation}
This fraction is shown as a function of column density in Fig~\ref{fig:f}.

\subsection{Spectra}
\label{sec:spec}

\begin{figure}
  \centering
  \includegraphics[width=\columnwidth,clip=]{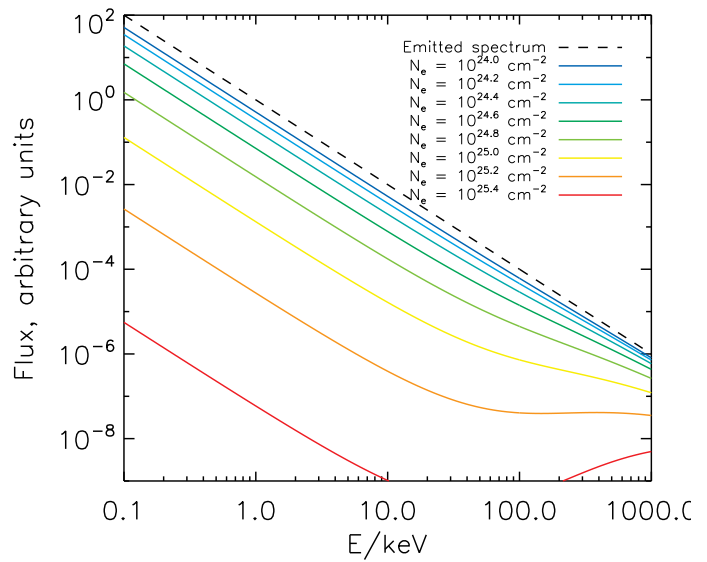}
  \caption{The model input and output spectra for the simulated GRB
           afterglow for a range of foreground column densities. }
  \label{fig:spectra}
 \end{figure}

\begin{figure}
\includegraphics[width=\columnwidth,clip=]{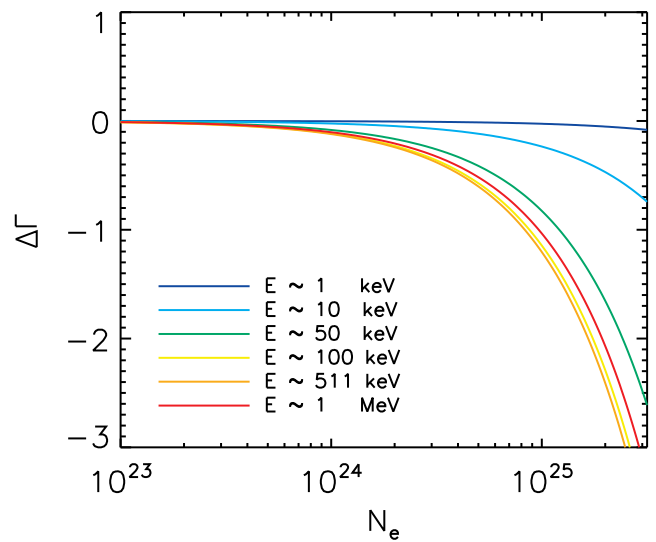}
\caption{Change $\Delta\Gamma$ in the observed spectral slope
         $\Gamma_{\mathrm{obs}}$ as a
         function of column density $N_e$, at various energies.}
\label{fig:Geff}
\end{figure}

Due to the decreasing
Compton cross section with photon energy, an emitted spectrum of constant
$\Gamma$ will have its high-energy end transmitted more efficiently than its
low-energy end, resulting in a harder slope at high energies
(Fig.~\ref{fig:spectra}).
In practice, this means that the spectral slope
will change as a function of energy.
Fig.~\ref{fig:Geff} shows how the observed spectral slope $\Gamma_{\mathrm{obs}}$
changes with optical depth.
Assuming an intrinsically energy-independent spectral slope $\Gamma$, if a
broad energy range is observed the change in $\Gamma_{\mathrm{obs}}$ as a
function of energy reveals the optical depth of the intervening cloud.  For
example, for $\Gamma = 2$ a column density of $N_e \simeq
10^{25}$\,cm$^{-2}$ will result in
an observed slope $\Gamma_{\mathrm{obs}} \lesssim 1$ at high energies.

\section{Discussion}
\label{sec:discussion}

To determine limits on the Compton depth of a given burst we have three
diagnostics at our disposal: the lightcurve, the total luminosity, and the
spectrum.  We have shown that the flux in the lightcurve (and the fluence
over the observed timescales of GRBs), is dominated by the directly
transmitted component even up to high column densities
($\gtrsim3\times10^{25}$\,cm$^{-2}$, see Fig.~\ref{fig:lightcurve}).  This
means that we cannot use the smearing of the shape of the lightcurve to
discern even moderately high column density bursts.  

The limits that can be placed on the column density based on the luminosity
of a burst are not very precise because we do not know intrinsically what
the apparent luminosity of a burst was.  We do, however, have a distribution
of equivalent isotropic luminosities of GRBs, and we know that bursts with
luminosities above $\sim2\times10^{54}$\,erg are rarely detected
\citep{2008MNRAS.387..319G}.  If we assume that there is no bias relating
luminous bursts with high column density sightlines, it suggests that any
burst is likely to be intrinsically less luminous than this value.  We can
therefore use the luminosity of a burst to determine the diminution in flux
possibly attributable to Compton scattering.  In Fig.~\ref{fig:f} we show
the fraction of emitted photons in the prompt emission as a function of
column density.  In the general case at high redshift, we are unlikely to
detect any burst more than two orders of magnitude fainter than
$10^{54}$\,erg with \emph{Swift}-BAT, and so the column density
for high-redshift \emph{Swift} bursts must always be
$\lesssim10^{25}$\,cm$^{-2}$ (see Fig.~\ref{fig:f}).  In the specific case
of the $z=6.3$ GRB\,050904, this would imply a maximum column density of
only a few times $10^{24}$\,cm$^{-2}$ because of the high apparent
luminosity of this burst.  However, individual cases are always vulnerable
to the argument that they may have an exceptional intrinsic luminosity. 
Even in the general case, it is possible that $z>6$ GRBs may be inherently
more luminous than lower redshift GRBs, weakening this argument further.  It
might be possible to predict the apparent luminosity of a burst based on a
number of correlations of burst properties with luminosity, the most
well-regarded of which is the peak energy--isotropic luminosity correlation
\citep[the ``Amati relation'', see][and references
therein]{2008MNRAS.387..319G}.  However, at the moment none of these
correlations is very tight and their general validity is controversial.  We
therefore turn to the spectra to examine whether they can provide more
robust constraints.

The spectra show a clear deviation from a power-law at column densities
above $\sim10^{25}$\,cm$^{-2}$, if we have simultaneous broadband coverage
from a few keV to a few hundred keV.  Currently, we do not typically obtain
simultaneous soft and hard X-ray detections of a GRB, though there are a few
exceptions where the burst was long and the \emph{Swift} response time
short.  A good example is, in fact, GRB\,050904 where the spectrum of the
late prompt phase is detected in the range 2--700\,keV in the rest frame
\citep{2006ApJ...637L..69W,2007A&A...462...73C}.

In general, at $10^{25}$\,cm$^{-2}$, the spectral slope has changed by 1 in
the 0.1--1\,MeV range (Fig.~\ref{fig:Geff}).  Such a strong change in
the slope of a burst would be readily discernible in late prompt data with
simultaneous X- and gamma-ray spectra, and it seems likely that this
spectral characteristic represents the strongest constraint on the
column densities.  It should be cautioned that in most GRBs strong intrinsic
spectral evolution of the burst occurs, causing the low energy data to
change spectral slope substantially.  In the case of GRB\,050904, the slope
changes in the first few minutes from $\Gamma\sim1.2$ to $\Gamma\sim1.9$. 
Such early spectral evolution is common in GRBs.  It is therefore essential
that in looking for this effect the high and low energy data be
simultaneous.  Using GRB\,050904, we note that the BAT and XRT
contemporaneous spectra have compatible power-law slopes, with uncertainties
of $\sim0.1$ on their spectral indices \citep{2007A&A...462...73C}. 
Disregarding the relatively small flux and cross-calibration uncertainties,
this similarity of the observed spectral slopes between the XRT and BAT
yields a $3\sigma$ limit on the column density of
$\sim5\times10^{24}$\,cm$^{-2}$.  This is a tight constraint, but still
three times as high as the $\tau=1$ limit imposed by
\citep{2007ApJ...654L..17C} and yields therefore a metallicity limit of
$>1$\% of the solar value (following their analysis of the X-ray
absorption). It is worth noting that the detection of the high column density in
GRB\,050904 is disputed \citep{2007ApJ...663..407B}, and there is the risk
of confusing intrinsic spectral curvature for photoelectric absorption where
we do not have high signal-to-noise ratio data at multiple epochs.

\subsection{Population~III progenitors}
At very high redshifts, GRBs may be formed from primordial stars --
population~III stars.  If a GRB occurs at very high redshift, a
discriminator between a population~III GRB and a non-primordial GRB would be
useful.  At $z=9-13$, for example, the lowest feasible detection of
photoelectric absorption is $\sim10^{23}$\,cm$^{-2}$ in units of equivalent
hydrogen column density at solar metallicity.  Assuming a similar limit on
the total electron column density as found for GRB\,050904, of
$\sim5\times10^{24}$\,cm$^{-2}$, where soft X-ray absorption is detected,
this would imply a metallicity limit at least $\gtrsim0.02\,Z/Z_\odot$ in
the $\sim1$\,pc environment of the GRB.  This should be sufficient to
exclude a population III star as the GRB progenitor.

\section{Conclusions}

We have examined the prospect of using only the high-energy emission of GRBs
to place a lower limit on metallicities around them; the soft X-ray
photoelectric absorption providing a measure of the total metal column
density and the Compton thickness limit a maximum column density of
electrons and hence hydrogen.  We find that smearing of the lightcurves does
not provide strong constraints on the column densities, while apparent
luminosity and deviations from a power-law spectral shape provide stronger
constraints.  The spectral constraints are more reliable, but require the
assumption of a single power-law spectral shape and simultaneous gamma- and
X-ray coverage.  The apparent luminosity constraints are more readily
applicable, but are less certain, and rely heavily on an assumption that we
know the distribution of apparent luminosities of GRBs, which is especially
uncertain at high redshift.  We find for the individual case of GRB\,050904,
the best limit that can be obtained from spectra corresponds to
$\sim5\times10^{24}$\,cm$^{-2}$, three times what was previously assumed. 
While the results are not very constraining for most GRBs, with bright,
long-lasting bursts, with detections of very high metal column densities,
the method could be used to exclude a population III progenitor for a high
redshift burst.

\begin{acknowledgements}
The Dark Cosmology Centre is funded by the DNRF.  PL acknowledges funding
from the Villum Foundation.  The simulations were performed on the
facilities provided by the Danish Center for Scientific Computing.  We would
like to thank Jens Hjorth and Anja C. Andersen for a critical reading of
the manuscript
\end{acknowledgements}

\bibliography{apj-jour,grbs}

\end{document}